\documentstyle[preprint,aps,prb,epsf]{revtex}
\begin{document}
\draft
\title{
Electronic Raman scattering and photoluminescence from  
La$_{0.7}$Sr$_{0.3}$MnO$_3$ exhibiting giant magnetoresistance 
}
\author{
Rajeev Gupta, A. K. Sood\cite{byline}, R. Mahesh$^{*}$ and 
C. N. R. Rao$^{*}$\\
}
\address{ 
Department of Physics\\
Indian Institute of Science\\
Bangalore, India 560 012
}
\address{$^{*}$
Solid State and Structural Chemistry Unit\\
Indian Institute of Science\\
Bangalore, India 560 012
}
\maketitle
\date{\today}

%\widetext

\begin{abstract}
Raman and Photoluminescence (PL)  experiments on correlated metallic 
La$_{0.7}$Sr$_{0.3}$MnO$_{3}$ have been carried out using different
excitation wavelengths as a function of temperature from 15 K to 300
K. Our data suggest a Raman 
mode centered at 1800 cm$^{-1}$ and a PL band at 2.2 eV. The intensities of
the two peaks decrease with increasing temperature. 
The Raman mode can be attributed to a plasmon excitation whose frequency
and linewidths are consistent with the measured resistivities. The
PL involves intersite electronic transitions of the manganese ions.
\end{abstract} 


%\narrowtext

The recent observations of colossal magnetoresistance has stimulated a renewed
interest in the electronic properties of doped
La$_{1-x}$A$_{x}$MnO$_{3-\delta}$ (A = Sr,Ba,Ca,Pb and
vacancies) and other transition metal oxides having strong
electron correlations \cite{jin}. The rich phase diagram of
La$_{1-x}$Sr$_{x}$MnO$_{3}$ shows a variety of phases like
paramagnetic insulator, ferromagnetic metal, paramagnetic
metal, spin canted insulator and ferromagnetic insulator as
a function of doping $x$ and temperature. For 0 $< x < $ 0.2, the
materials are insulating at all temperatures and are
antiferromagnetic or ferrimagnetic at low temperatures. 
In the range 0.2 $< x <$ 0.5 the system shows a temperature induced
transition at T$_{c}$($x$) from the ferromagnetic metal ( at $T < T_c$ ) to
a paramagnetic insulator. The end member ($x$
= 0) is a charge transfer antiferromagnetic insulator having gap
corresponding to the charge transfer excitation from oxygen 2p to
manganese 3d state. Out of the (4-$x$) manganese d electrons, three
electrons occupy the tightly bound $d_{xy}$, $d_{yz}$ and $d_{xz}$
orbitals with very little hybridization with the oxygen 2p states and can be
considered as a local spin S$_c$ of 3/2. The remaining (1-$x$) electrons
occupy the $e_g$ state made of the $d_{x^2-y^2}$ and $d_{z^2}$
orbitals and are strongly hybridised. There is a strong exchange
interaction $J_H$ (Hund's coupling) between the 3d $t_{2g}$ local
spin and the 3d $e_g$ conduction electron. The $e_g$ level is further
split into $e_{g}^{1\uparrow}$ and $e_{g}^{2\uparrow}$ due to the
Jahn-Teller (JT) effect. The 
$t_{2g}^\uparrow$ -- $e_{g}^{2\uparrow}$
and $e_{g}^{1\uparrow}$ -- $e_{g}^{1\downarrow}$ separation is about 2
eV as calculated by Satpathy et al using the local spin density
approximation \cite{satp}. The estimates of JT split $e_g$ band namely 
$e_{g}^{1\uparrow}$ -- $e_{g}^{2\uparrow}$ = 4$E_o$ using the oxygen
bond stretching frequencies to be 560 cm$^{-1}$ is 2.4 eV. It can,
however, 
range from 0.4 eV to 4 eV as argued by Millis \cite{mill1}. 

The metal -- insulator transition in the intermediate doping range is
qualitatively understood using the Zener's "Double Exchange" (DEX)
model, in which the e$_g$ electron hopping from site i to j must go
with its spin parallel to S$_{c}^{i}$ to its spin parallel to
S$_{c}^{j}$.  
Millis et al \cite{mill2} have shown that DEX alone
cannot explain many
aspects like the low
transition temperature $T_c$ and the large resistivity of $T > T_c$
phase or the sudden drop in 
resistivity below $T_c$. 
They have proposed that in addition to DEX, there is a strong electron
phonon coupling such that the slowly fluctuating local Jahn-Teller
distortions localise the conduction band electrons as polarons.  
As temperature is lowered the effective hopping matrix element $t_{eff}$
characterising the electron itineracy increases and the ratio of JT
self trapping energy $E_{JT}$ to $t_{eff}$ decreases. The JT
distortion has to be dynamic because a
static JT effect would cause a substantial distortion of the
structure and the material would be antiferromagnetic.
Coey et al\cite{coey} have argued from the experimental magnetoresistance data
that for T $<$ T$_c$,the e$_g$ electrons are delocalised on an atomic
scale but the spatial fluctuations in the Coulomb and spin 
dependent potentials
tend to localise the $e_g$ electrons in wave packets larger than the
Mn -- Mn distance. It has also been noted \cite{coey,mahe1} that doped
manganites are unusual metals having resistivities greater than the
maximum Mott resistivity (1 to 10 mohm cm) and a very low density of
states at the Fermi level \cite{sarm}.

Optical conductivity measurements on La$_{0.825}$Sr$_{0.175}$MnO$_3$
as a function of temperature by Tokura et al\cite{okim} in the range of 0
-- 10 eV show a band at $\sim$ 1.5 eV and spectral weight is
transferred from this band to low energies with decreasing
temperature. This band at $\sim$ 1.5 eV has been interpreted due to the
interband transitions between the exchange split spin polarized $e_g$
bands. At $T < T_c$, the conductivity spectrum is 
dominated by intraband transitions in the $e_g$ band. 
A similar feature at $\sim$ 1 eV has been observed in
Nd$_{0.7}$Sr$_{0.3}$MnO$_3$ which shifts to lower energies with
decreasing temperature which is argued to be consistent by taking
into account the dynamic JT effect\cite{kapl}. There is no reported work on Raman 
scattering in
these systems. Millis\cite{mill1} has suggested that the transition between the
$e_g$ levels split by the JT interaction can be Raman active. The
electronic Raman scattering can be observed from the single particle
and collective plasmon excitations. The crystal
structure with space group R$\overline{3}$c ($D_{3d}^{6}$) 
with two formulas in the
unit cell has A$_{1g}$ and 4E$_g$ Raman active modes. Our
objective was to study vibrational and electronic Raman scattering in
doped manganites as a function of temperature.       
In this paper we report the Raman scattering 
from doped La$_{1-x}$Sr$_x$MnO$_3$ ($x$ = 0.3) from 15 K to 300 K
attributed to a plasmon mode and a photoluminescence band at $\sim$
2.2 eV.

Polycrystalline pellets of La$_{0.7}$Sr$_{0.3}$MnO$_3$ prepared by citrate
-- gel route and sintered at 1473 K with an average grain size of
$\sim$ 3.5 $\mu$m were used \cite{mahe1}.
The samples on which detailed studies were done have a resistivity of
0.4 m$\Omega$ cm at 15 K and 3.5 m$\Omega$ cm at 300 K. It shows a phase
transition from a ferromagnetic metal to a
paramagnetic insulator at T$_c \sim$ 380 K. The Raman measurements were
carried out in the spectral range of
200 cm$^{-1}$ to 6000 cm$^{-1}$, at different temperatures
from 15 K to 300 K. The spectra were recorded in the back
scattering geometry using Spex Ramalog with photon counting
detector (photomultiplier tube RCA C 31034  with GaAs cathode ) using
514.5 nm, 488 nm and 457.9 nm lines of an
Argon ion laser (power density of $\sim$ 1500 watts cm$^{-2}$ at the
sample). The pellet of thickness $\sim$ 2 mm were mounted on the copper
cold finger
of the closed cycle helium refrigerator (RMC model 22C
CRYODYNE) using thermally cycled GE ( M/s. General Electric, USA) varnish. The
temperature of the cold finger was measured using a
Platinum 100 sensor coupled to a home made temperature
controller. The temperatures quoted are those of the cold
finger and were measured to within a accuracy of $\sim$ 2
K. The experiments were done on three differently prepared pellets of
La$_{0.7}$Sr$_{0.3}$MnO$_{3}$ and the results were similar to the
ones reported here.

Fig. 1 shows the recorded spectra at 15 K for
the three different excitation wavelengths $\lambda_L$ of
514.5, 488 and 457.9 nm. Fig.
2,3 and 4 show the spectra
at different
temperatures for various excitation wavelengths used. 
It is clear from the spectrum recorded using 514.5 nm as shown in
Fig. 1, that the observed lineshape has two contributions.
The spectra were, therefore, least
square fitted to
a sum of two Lorentzians shown by the thick solid line along with
the data (open circles).The individual components are shown by thin
solid lines.The intensities of the two peaks 
decrease with the increase in temperature. Since the spectra are
rather weak and broad, the fitted parameters
of the Lorentzian function (intensity and linewidths) have large
error bars and hence their quantitative temperature dependence has
not been possible to determine. 

We have not been able to observe Raman scattering from the phonons. 
This can perhaps be due to the fact that the deviations from the cubic
structure (for which there are no Raman active modes) are rather
small. In collision dominated regime, the Raman scattering line shape
from the single particle electronic excitations is given by
I($\omega$) $\propto$ ($n(\omega,T) + 1)\omega\gamma$ B /($\omega^2 +
\gamma^2)$, where $(n(\omega,T) + 1)$ is the usual Bose Einstein
factor, B is the strength of scattering and $\gamma$ is the
relaxation rate \cite{ipat}. This Raman scattering has been observed in
La$_{1-x}$Sr$_{x}$TiO$_3$ by Tokura et al \cite{kats}. We have
not been able to resolve the contribution of the electronic Raman
scattering from the Rayleigh wing, presumably due to the poor
surface quality of our sintered pellets.

The spectra shown in Figs 1 and 2 recorded with $\lambda_L$ = 514.5
nm are a sum of two Lorentzians centered
at $\sim$ 950 cm$^{-1}$ and $\sim$ 1800 cm$^{-1}$. 
The spectra recorded with three
different excitation wavelengths are consistent with the
interpretation that the observed lineshape is a sum of Raman
mode centered at $\omega \sim$ 1800 cm$^{-1}$ and a PL band centred
at photon energy of 2.2 eV.
The latter will appear as an apparent Raman shift which varies
with different excitation wavelengths. 
The likely candidate for the Raman mode at 1800 cm$^{-1}$ can be a
collective electronic excitation namely the plasmon. Taking the excitation
to be a plasmon with frequency $\omega_p$ = 1800 cm$^{-1}$, the
number density of charge
carriers of effective mass $m^{\star}m_e$ can be obtained using the
expression 
$$ \omega_{p}^2 = 4\pi n_{p}e^{2}/(\epsilon_{\infty}m^{\star}m_e) ,$$  
where $\epsilon_{\infty}$ is the high frequency dielectric constant.
Taking $\epsilon_{\infty}$ to be same\cite{arim} as that of LaMnO$_3$,
$\epsilon_{\infty}$ = 4.9 we get $n_{p}/m^{\star}$ = 1.8 x 10$^{20}$
cm$^{-3}$, The number density calculated from the doping of
0.3 carriers per unit cell of cell volume\cite{mahe1} 64 $\AA^3$ is $n_d$ = 5 x
10$^{21}$ cm$^{-3}$. Taking the effective mass of the carriers to be 
$m^{\star}$ = 1, it is seen that $n_p$ = 0.04 $n_d$ i.e the 
actual number of carriers
is less than (1-$x$) per manganese site. This can be related to the
model of Coey et al\cite{coey} wherein the e$_g$ electrons though
delocalised on atomic scale, are in magnetically localised
wavepackets spread over the Mn -- Mn separation. If the localisation
energy of some carriers is less than $h\omega_p$, they will not
participate in the collective plasmon excitation. The localisation
need not be as magnetic polarons but may involve lattice polarons as
envisaged by Millis\cite{mill1,mill2}. The reduced number of
carriers is consistent with the Hall
measurements of Hundley as referred by Roder et al\cite{roder}. The
observed low density of states at Fermi energy\cite{sarm} also
corroborates the less number of carriers deduced from $\omega_p$. Perhaps
$m^{\star}$ can be greater than 1 which will reduce the difference
between the $n_p$ and $n_d$. The least square fit with the
Lorentzians yields the full width at half maximum $\Gamma$ of the Raman mode
to be $\sim$ 1400 cm$^{-1}$. The DC resistivity in the free electron model
is $\rho$ = $m / n_{p}e^{2}\tau$ ; where $\tau$ is the relaxation
time of the carriers. Putting $\tau^{-1}$ = $\Gamma$, $\rho$ can be
expressed in terms of $\omega_{p}^2$ and $\Gamma$ as  
$\rho$ = $4\pi\Gamma / (\epsilon_{\infty} \omega_{p}^2$).
Taking $\Gamma$ = 1400 cm$^{-1}$ and $\omega_p$ = 1800 cm$^{-1}$, we
get $\rho$ = 5.3 mohm -- cm, which is remarkably close to the
measured values and therefore gives confidence in our assignment of
the Raman mode to be due to the plasmon. The $\rho (T)$ and hence
$\Gamma$ increases with temperature and can therefore result in
substantial reduction of Raman intensities, as seen in our
experiments. The observed Raman frequency is much smaller than the
estimated 4E$_o$ and hence is not likely to be associated with the
transition between the JT split e$_g$ bands. 

Another interesting
aspect of the data shown in Figs 2 -- 4 is that the PL peak is also seen 
only in the low 
temperature metallic phase.
The photoluminescence band is therefore linked with the excitations
involving e$_g$ states at different manganese sites. The possible
transitions could be between the 
$t_{2g}^{\uparrow}$ -- $e_{g}^{2\uparrow}$, 
$e_{g}^{1\uparrow}$ --$e_{g}^{1\downarrow}$ or $e_{g}^{1\uparrow}$
--$e_{g}^{2\uparrow}$ states. It may be noted that the on-site transitions 
involving the above states 
are not optically dipole allowed
as per the selection rules but these can be allowed for the states on
different manganese sites. Since there is a strong temperature
dependence, we rule out the possibility that the PL
is due to transitions from the oxygen 2p states to the $e_g$ states. 
It is possible to have a transition from $e_{g}^{1\uparrow}$
(Mn$^{+3}$) to the unsplit $e_g$ level (Mn$^{4+}$) level on an adjacent
site. In contrast to Tokura's interpretation, the peak in the 
optical conductivity at $\sim$ 1 eV 
in La$_{0.825}$Sr$_{0.175}$MnO$_3$ and Nd$_{0.7}$Sr$_{0.3}$MnO$_3$ 
has been attributed by Millis et al\cite{mill1}
to the transition from e$_{g}^{1\uparrow}$ of Mn$^{3+}$ site to the
unsplit e$_g$ level of adjacent Mn$^{4+}$ site.
If this assignment is correct,
then the observed PL band at 2.2 eV in our experiments cannot be due
to the above transtion as the
energy gap is too small ($\sim$ 1 eV).
As the temperature is increased the electrons tend to get
localised\cite{mill2} and thus making it difficult to have intersite transitions.
The energy separations of the transitions mentioned before have not
been 
determined so far. We can get  rough estimates of these from the available
band structure calculations, with the caution that the one electron
band theory is not the correct picture for the strongly correlated
electron systems and the local spin density approximation
calculations underestimate the 
gaps.

Further experiments on these materials especially on single crytals
can provide more clues as to nature and symmetry of these excitations.
The Raman mode attributed to the plasmon should occur only in polarized
scattering configuration and single particle excitation contributions
can be unambigously distinguished from the Rayleigh wing.

We thank Prof. T.V. Ramakrishnan, R. Mahendiran
,Prof. A.K. Raychaudhuri and Prof. B.S. Shastry for useful
discussions. AKS thanks
Department of Science and Technology and RG thanks 
CSIR for the financial assistance.

%\end{document}

\begin{figure}
\caption{Raman Spectra recorded with different excitation wavelengths
514.5nm, 488 nm and 457.9 nm at 15 K. The solid lines shows the fitted
function (a sum of two Lorentzians) to the data shown by open
circles. The individual components are shown by thin solid lines.
The Raman and PL component are labeled as R and L, respectively.}
\label{Fig.1}
\end{figure}

\begin{figure}
\caption{Raman Spectra in the temperature range 15 K to 300 K
using 514.5 nm excitation wavelength. Fitted function of two
Lorentzians is shown by the
solid line for each spectrum. The intensities of the two modes decrease as
temperature increases.}
\label{Fig.2}
\end{figure}

\begin{figure}
\caption{Same as Fig. 2, but with different excitation wavelength of 488 nm.
}
\label{Fig.3}
\end{figure}

\begin{figure}
\caption{Same as Fig. 2, but with different excitation wavelength of
457.9 nm. 
}
\label{Fig.4}
\end{figure}

\end{document}